\documentclass[12pt,preprint]{aastex}  %>>> use for US letter paper

\newcommand {\ms}{\rm ms^{-1}}
\newcommand {\ET}{{\it ET}}

\shorttitle{First Planet Confirmation with \ET }
\shortauthors{van Eyken et al.}

\begin{document}

\title{First Planet Confirmation with a Dispersed Fixed-Delay Interferometer}

\author{J. C. van Eyken\altaffilmark{1}, J. Ge\altaffilmark{1}, S. Mahadevan\altaffilmark{1}, C. DeWitt\altaffilmark{1}}
\affil{Dept. of Astronomy \& Astrophysics, Penn State University, 525
Davey Laboratory, University Park, PA 16802
%\email{vaneyken@astro.psu.edu, jian@astro.psu.edu,
%suvrath@astro.psu.edu, dewitt@astro.psu.edu}
}
\altaffiltext{1}{Visiting Astronomer, Kitt Peak National Observatory, National Optical Astronomy Observatory, which is operated by the Association of Universities for Research in Astronomy, Inc. (AURA) under cooperative agreement with the National Science Foundation. E-mails: vaneyken@astro.psu.edu, jian@astro.psu.edu, suvrath@astro.psu.edu, dewitt@astro.psu.edu.}

\begin{abstract}
The {\it Exoplanet Tracker} is a prototype of a new type of
fibre-fed instrument for performing high precision relative
Doppler measurements to detect extra-solar planets. A combination
of Michelson interferometer and medium resolution spectrograph,
this low-cost instrument facilitates radial velocity measurements
with high throughput over a  small bandwidth ($\sim300$\AA), and
has the potential to be designed for multi-object operation with
moderate bandwidths ($\sim$ 1000\AA). We present the first planet
detection with this new type of instrument, a successful
confirmation of the well established planetary companion to 51
Peg, showing an rms precision of $11.5\ms$ over five days. We also
show comparison measurements of the radial velocity stable star,
$\eta$ Cas, showing an rms precision of $7.9\ms$ over seven days.
These new results are starting to approach the precision levels
obtained with traditional radial velocity techniques based on cross-dispersed echelles. We
anticipate that this new technique could have an important impact
in the search for extra-solar planets.
\end{abstract}


%------------------------------------------------------------------

\section{INTRODUCTION}
Of the more than one hundred extra-solar planets that have been found to date,
the vast majority have been found using the radial velocity (RV)
technique (e.g. \citet{Butler1996, Baranne1996}). Current
approaches to making RV measurements rely on using very high
resolution echelle spectrographs, employing cross-correlation or fits to line profiles in
stellar spectra to determine Doppler shifts in the centroids of
the lines. While the RV technique has been the most successful
technique for locating extra-solar planets, traditional echelles
have relatively low light throughput, large instrument volume, and
tend to be very expensive. In addition, they cover only a single object
in each observation. The low light throughput limits survey
sensitivity to relatively bright stars and single object operation
leads to slow survey speeds.

The {\it Exoplanet Tracker} (\ET) is a prototype of a new type of
fibre-fed RV instrument based on a dispersed fixed-delay
interferometer, a combination of a Michelson interferometer
followed by a low or medium resolution post-disperser. This
combination has been suggested for spectroscopic applications as
early as the 1890's \citep{EdserButler}, and was proposed for
precision Doppler planet searches by D. J. Erskine in 1997
\citep{ErskineGe, GeErskine}; a similar approach is discussed in \citet{Mosser2003}. The effective resolution of the
instrument is determined primarily by the interferometer, so the
post-dispersing spectrograph can be of much lower resolution than
in traditional techniques, and consequently can have much higher
throughput \citep{GeTheory, Ge2003a, Ge2003b}.

The cost of the instrument is comparitively low, and furthermore it operates
in a single-order mode: a single spectrum only takes up one strip
along the CCD detector. Spectra from multiple stars can be lined up
at once on a single detector to increase survey speed \citep{GeTheory}.
In combination with a wide field multi-fibre telescope,
multi-object surveying should therefore be achievable
\citep{SuvrathSPIE}, with the potential to rapidly increase the number of
known extra-solar planets.

%Our current understanding of extra-solar planets is limited by the
%small sample size available. With the ability to search faint
%stars with very high survey speed, it should be possible to
%rapidly increase the number of known extra-solar planets.

The instrument works by producing a long-slit stellar spectrum
`channeled' with fringes, also known as Edser-Butler fringes
\citep{EdserButler, SummerSchool, GeTheory}. Sinusoidal
interference fringes are formed along the slit direction wherever
there are spectral lines. Doppler shifts in the underlying
spectrum result in directly proportionate phase shifts in these
fringes. Hence, we measure shifts of the sinusoids in the slit
direction, rather than shifts of the spectrum itself in the
dispersion direction as in traditional techniques. By fitting sine
functions to the CCD response along each wavelength channel on the
detector and combining the results from all channels, we are able
to accurately measure any changes in the Doppler shift of an
object. The concept is described in more detail in
\citet{GeTheory, vaneykenSPIE}.

In this paper, we report Doppler RV curves of the known planet-bearing
star, 51 Peg \citep{51PegDiscovery}, and a RV stable star, $\eta$ Cas with \ET~at the KPNO
2.1m telescope. This represents the first
planet detection using this independent new technique.

%------------------------------------------------------------------

\section{OBSERVATIONS}

The observations were conducted with a prototype of \ET~during an
engineering run at the KPNO 2.1m telescope in August 2002, built
largely from cheap off-the-shelf components. The observations and details of the
instrument setup were reported in \citet{Ge2003a}.

The spectrograph
operating resolution was measured at $R\approx4540$. Using a KPNO
1k$\times$3k back-illuminated CCD gave a wavelength coverage of
$\sim270$\AA\ centred around 5445\AA. The image was spread over
$\sim300$ pixels in the slit direction, giving a total of around
12 periods of fringing. An iodine vapour cell was inserted into the
beam as a Doppler zero velocity reference, with its temperature
stabilised to $60\pm0.1^\circ{\rm C}.$

During the run, we were able to obtain regular observations of a
number of stars, including known planet bearing stars 51 Peg,
$\upsilon$ And and HD209458; RV stable stars $\eta$ Cas, $\tau$
Ceti and and 31 Aql; and a bright star, $\alpha$ Boo, over a
period of about seven days \citep{Ge2003a}. %The weather was good for
%the most part, with typical seeing of around 1.7 arc-sec.
In this
paper, only results from 51 Peg and $\eta$ Cas are reported.

%----------------------------------------------------------------
\section{DATA ANALYSIS}

Raw spectra were first trimmed and dark subtracted using standard
{\it IRAF} routines, with bias being subtracted along with the
darks in one step. Pixel-pixel flatfielding was performed using
non-fringing quartz-lamp continuum spectra as flatfields, where
the fringes were eliminated by rapidly oscillating the interferometer PZT mirror during the exposure.

The rest of the data reduction was then performed using custom
software written in the {\it IDL} data analysis language, by {\it
Research Systems Inc.} Images were `self illumination corrected' using an algorithm to
extract the underlying continuum illumination function from each
image, which is divided out. This avoids problems with changes in
the illumination over time. The spectra were then corrected for
slant so that the slit direction was exactly aligned with the CCD
pixel axes. They were then low-pass Fourier filtered in order to
remove the interferometer comb, the series of parallel fringes
that would be present if pure white light were to be observed and
which contains no Doppler information itself.

After these pre-processing steps, the phase and visibility were determined for each wavelength channel by fitting a sin wave to each
column of the CCD image, each pixel being weighted according to
the number of counts in the original non-flatfielded data on the
assumption of photon noise dominated error. Since fringe spatial
frequency varies only slowly as a function of wavelength, we fit a
smooth function to the frequencies obtained from the
sinusoid fits, and then performed a second pass with the frequencies
fixed to match this function, helping to reduce random errors.

To a good approximation, the combined iodine/star data frames can
be considered a linear summation of the complex visibilities of
the individual iodine and stellar spectra (where complex
visibility is defined as ${\bf V} = Ve^{i\phi}$, with $V$ the
fringe visibility and $\phi$ the phase offset). Pure stellar and
pure iodine template spectra were taken at the beginning of each
observation, and these were used to mathematically extract the
phase shifts of the star and the iodine individually, and hence
calculate the intrinsic stellar velocity shift corrected for
instrumental drifts.

Finally, the RV due to the motion of the Earth was subtracted to
leave an intrinsic stellar relative velocity curve. Currently the
exposure time is taken to be the centre of the exposure (although
this is by no means necessarily ideal).

Error bars are based on the standard statistical curve-fitting
errors determined during measurement of phase and visibility. The
errors are translated to error bars through calculations
appropriate to the algorithms used to extract the final intrinsic
stellar RV. They are expected to give a reasonable guide to the
random scatter expected in the data, although they may not catch
all systematic errors.

On closer inspection, the data were found to show
variation in the fringe phase and visibility along the
length of the slit. We therefore cut the spectra into three slices
along the dispersion direction and treated each slice separately,
in order to obtain sinusoidal fits less affected by this
systematic error. A weighted average of the three results was then obtained to give a final RV plot.

%%%%%%%%%%%%%%%%%%%%%%%%%%%%%%%%%%%%%%%%%%%%%%%%%%%%%%%%%

\section{RESULTS}

\begin{figure}
\plotone{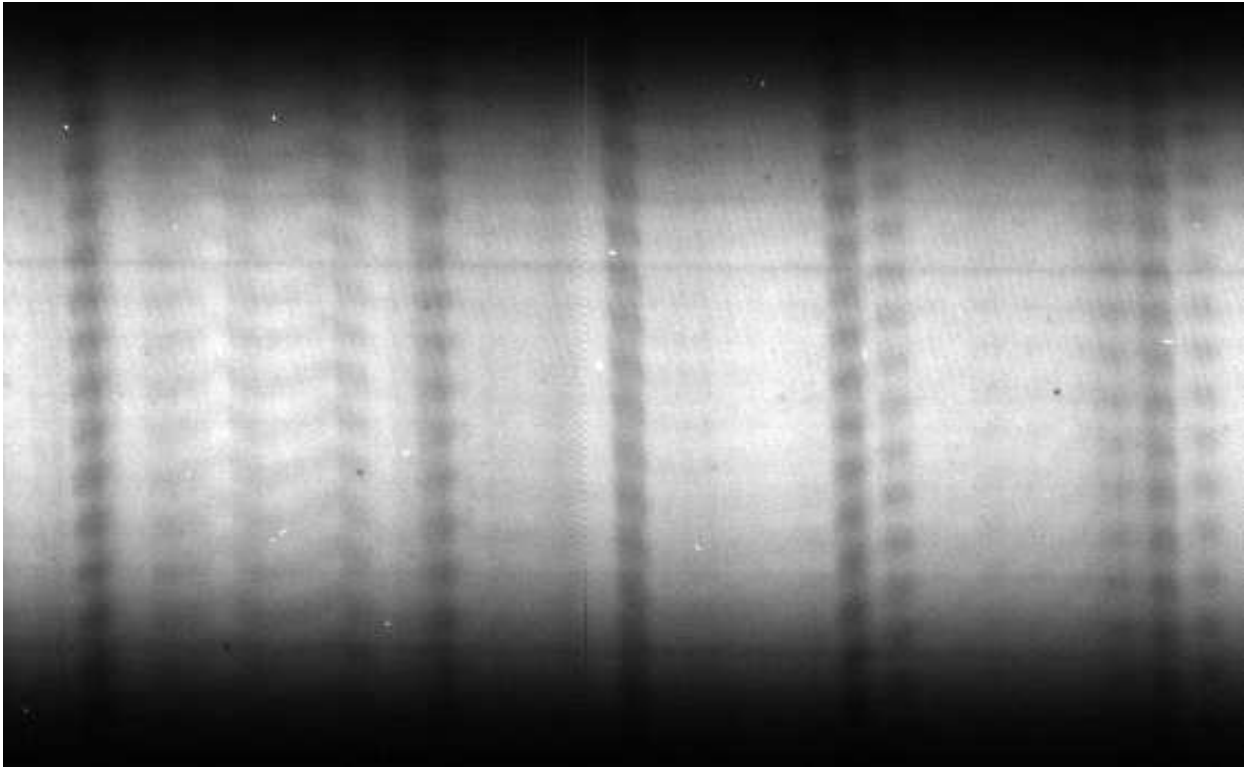} %, raw_51peg.eps, raw_hd209458.eps}
\caption{\label{fig:rawspectra} Raw fringing spectrum of 51 Peg
with iodine, obtained at KPNO on the night of 2002 Aug 14. (mag
5.5, S/N$\sim50$ per pixel)}
\end{figure}

Part of the raw fringing spectrum for 51 Peg with the overlayed
iodine spectrum is shown in figure \ref{fig:rawspectra}, obtained
in 25 minutes at visual magnitude 5.5 with S/N per pixel in the
central strip of around 50. Typical exposure times for $\eta$ Cas
(mag 3.5) were 30 min at an S/N of 80 (including iodine cell losses).

\begin{figure}
\plotone{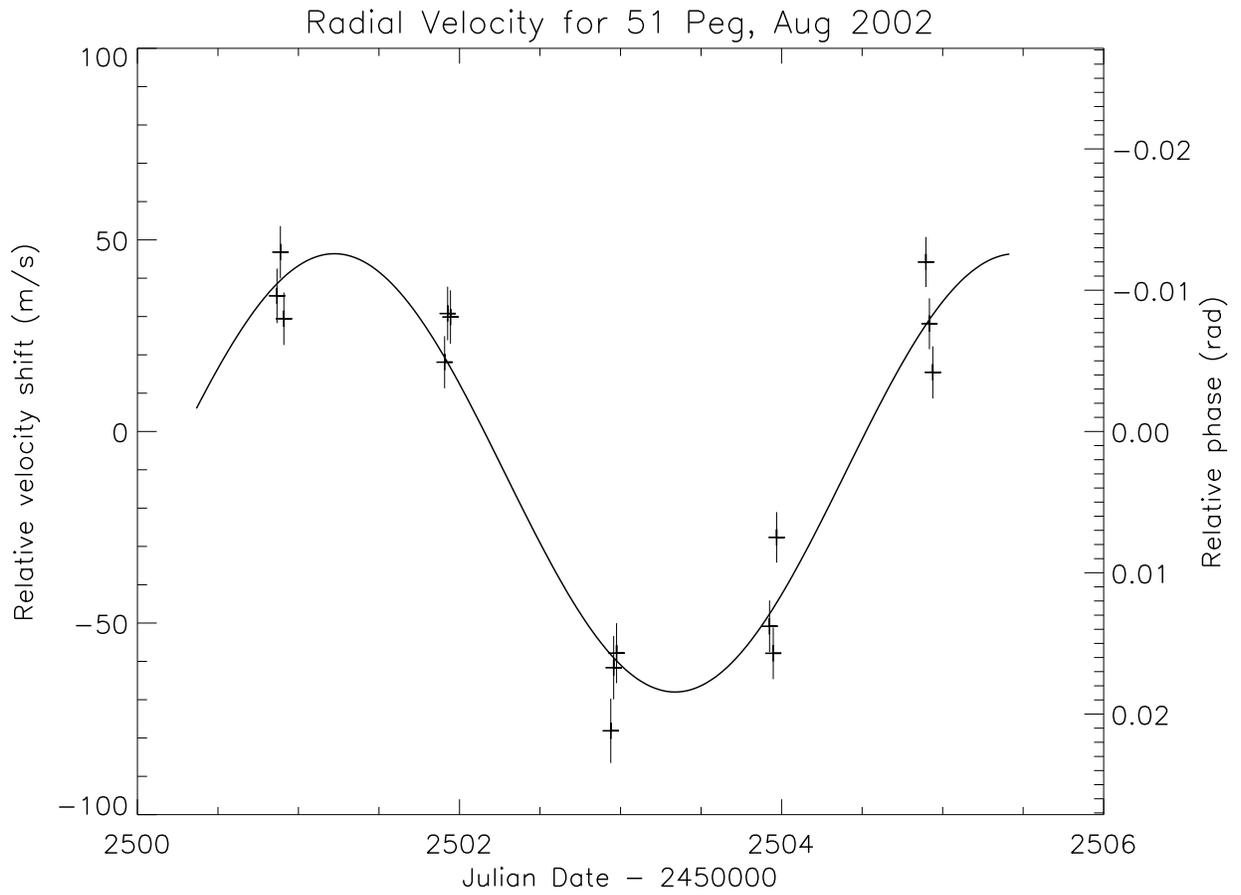}
\caption{\label{fig:51Peg} Radial velocity measurements for 51
Peg, with the predicted curve over-plotted. RMS residuals
are $11.5\ms$.}
\end{figure}

%\subsection{51 Peg B}
Figure \ref{fig:51Peg} shows the radial velocity variation
measured for 51 Peg after diurnal motion is subtracted.
The zero point is chosen arbitrarily. Over-plotted is the expected
curve extrapolated from the most recently determined orbital
parameters \citep{51peg_orbit}. The same data are listed in table
\ref{tab:51peg}. S/N per pixel ratios obtained were in the range
$\sim$40--60 for star+iodine spectra. The templates used for the
processing are from the night of August 16 (August 17 UT), and S/N for
the iodine and star templates were approximately 300 and 70
per pixel respectively (for the central strip).
Averaging over the three detector strips gives an rms deviation
from the predicted curve
 of $11.5\ms$. The value of the reduced $\chi^2$ is
2.70.\footnote{These results represent a substantial improvement
over our previously reported measurements \citep{vaneykenSPIE},
due in part to using all three detector strips and also to several improvements in
the reduction software.}

%\subsection{$\eta$ Cas}

\begin{figure}
%%\epsscale{1.1}
\plotone{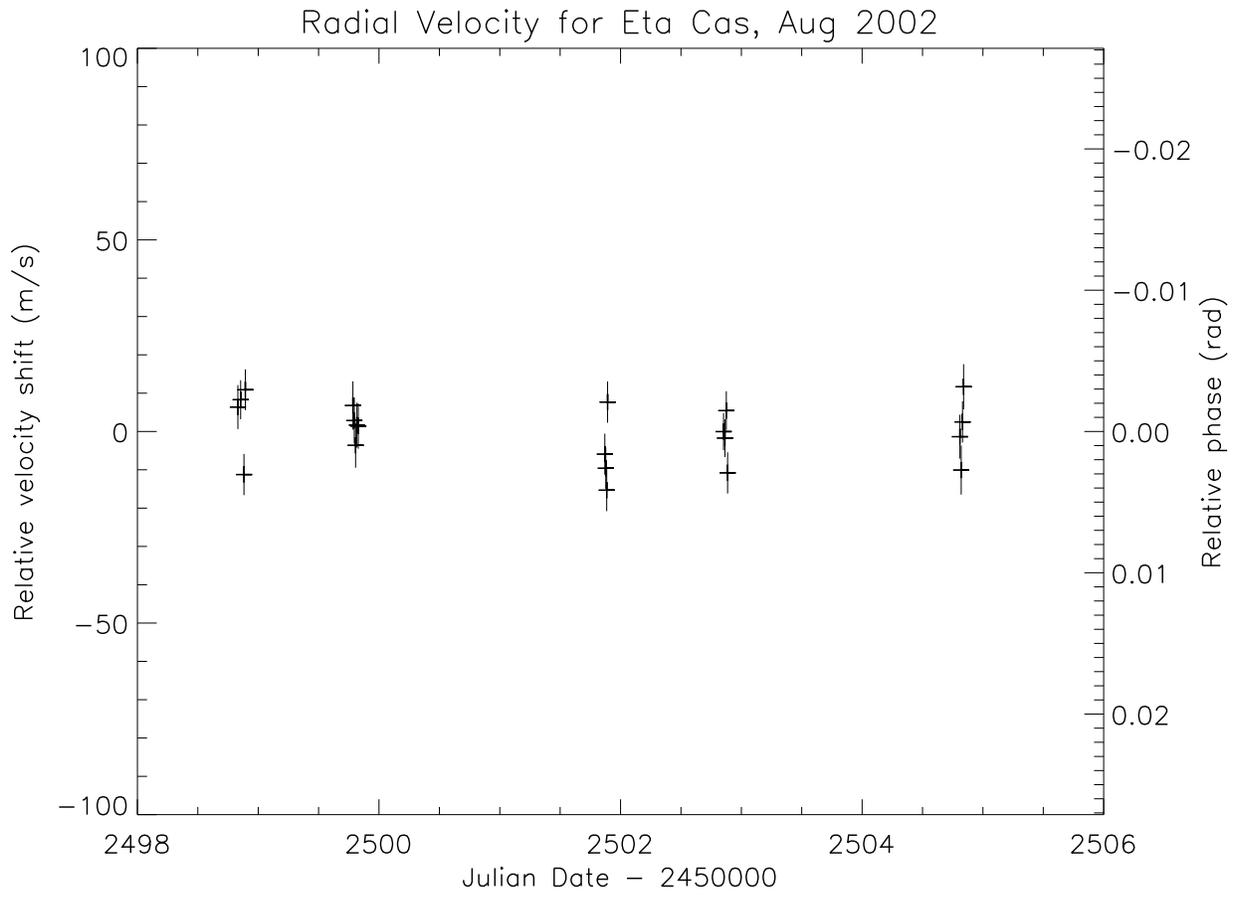}
\caption{\label{fig:EtaCas}  RV measurements for $\eta$ Cas, an RV
stable star, expected to show zero shift. RMS residuals are
$7.9\ms$.}
\end{figure}

Residuals after diurnal correction for the star $\eta$ Cas are
shown in figure \ref{fig:EtaCas} and table \ref{tab:EtaCas}, using
templates from the night of Aug 15 (Aug 16 UT). $\eta$ Cas is a
known RV stable star (W. D. Cochran 2002, private communication) and is
therefore expected to show zero shift at our current level of precision. The three image strips are
averaged, weighted according to flux. The rms scatter is $7.9\ms$,
with a reduced $\chi^2$ of 2.03. Typical S/N per pixel in the central
strip is around
70--90 for star+iodine spectra, 270 for the iodine template, and 100
for the star template.

%\subsection{Throughput}

Under 1.5 arc-sec seeing conditions, we obtained a total
instrument throughput of $\sim4\%$, from above the atmosphere to
the detector, including sky, telescope transmission, fibre loss,
instrument and iodine cell transmission, detector quantum
efficiency, and using only one interferometer output. Excluding
slit loss, the transmission of the instrument itself from fibre to
detector was 19\%.

%%%%%%%%%%%%%%%%%%%%%%%%%%%%%%%%%%%%%%%%%%%%%%%%%%%%%%%

\section{DISCUSSION}
\label{sec:discussion} It is possible to make an estimate of
the photon limited error for this instrument using the following
analysis. Following \citet{GeTheory}, the error $\sigma_i$ in
velocity from a single wavelength channel $i$ due to photon noise
alone can be calculated as a function of fringe visibility,
$\gamma_i$, and total photon flux in the channel, $F_i$. Using a slightly
more accurate derivation than given in \citet{GeTheory} gives the
relation:

\begin{equation}\label{eqn:jian} \sigma_i = \frac{c\lambda}{\pi d\gamma_i\sqrt{2F_i}},
\end{equation}

where $c$ is the speed of light, $\lambda$ is the operating
wavelength, and $d$ is the path difference between the
interferometer arms. The combined error over all 
channels is then given by:

\begin{equation}\label{eqn:individual} \sigma = [\sum_{i} 1/\sigma_i^2]^{-\onehalf}.
\end{equation}

This gives us an estimate of the error due to the photon noise in
one complete fringing spectrum. In order to estimate the photon
error for the final iodine reference corrected RV measurement, we
must combine the errors from the templates, $\sigma_{\rm I_2,
template}$ and $\sigma_{\rm star, template}$, and from the
combined star+iodine data. We treat the combined data as
consisting of two separate components, with errors $\sigma_{\rm
I_2, data}$ and $\sigma_{\rm star, data}$. The final relative
velocity measured, $V$, is given by $V = (V_{\rm star,
data}-V_{\rm star, template}) - (V_{\rm I_2, data} - V_{\rm I_2,
template}) $, and so the final error is obtained by quadrature
addition:

\begin{equation}\label{eqn:final} \sigma_V^2 = \sigma_{\rm star, data}^2 + \sigma_{\rm
star, template}^2 + \sigma_{\rm I_2, data}^2 + \sigma_{\rm I_2,
template}^2. \end{equation}

We find the errors due to the templates using equations \ref{eqn:jian} and \ref{eqn:individual}. To find the errors $\sigma_{\rm I_2, data}$
and $\sigma_{\rm star, data}$, we take the errors calculated for
the templates, and scale these to find the values that they should
have at the S/N of the data, noting from equation \ref{eqn:jian}
that the errors scale as the reciprocal of S/N (where
S/N$=\sqrt{F}$).

The results of these calculations are shown in table
\ref{tbl:errors}. We find final photon limiting precisions
(averaged over all data points) of $11.0\ms$ for 51 Peg and
$8.1\ms$ for $\eta$ Cas. Within the uncertainty in the rms
residual values obtained for the data due to the small number of
data points ($11.5\pm2.1\ms$ for 51Peg and $7.9\pm1.2\ms$ for
$\eta$ Cas), we find a good match with the data and conclude that
we have reached the photon limit: the reduction software has
successfully extracted the maximum possible information from the data. It is
important to note, however, that these values for the photon limit
are those expected {\em given} the fringe visibility that was
obtained. Various instrument effects (for example defocus) can
reduce the visibility from its optimum and hence reduce the
precision. It is therefore possible that the intrinsic limit is
somewhat lower for an ideally optimised instrument.

We note the large contribution to the errors due to the iodine
reference. Though the iodine can be measured to very high accuracy
for the template since a quartz lamp is used for illumination, the
iodine in the combined star/iodine images has much lower S/N, and
this becomes an important source of error. In both the 51 Peg and
the $\eta$ Cas cases, the error due to the iodine is comparable to
that of the star itself.

Given this photon limit, the error bars in the data appear to be
underestimated (leading to the large values for the reduced $\chi^2$).
A possible cause of this is the low pass Fourier filtering
that is done to remove the interferometer comb. In addition to
removing the comb, filtering has the effect of smoothing the
photon noise in the data, reducing the residuals in the sinusoid
fits to the fringes and thereby reducing the resulting error
estimates for each fringe. This may be an artificial
effect, however, which in fact does not improve the precision of
the fits. The
extent to which this effect occurs is under investigation.

%%%%%%%%%%%%%%%%%%%%%%%%%%%%%%%%%%%%%%%%%%%%%%%%%%%%%%%%

\section{CONCLUSION}

We have achieved $11.5\ms$ RV precision over five days of
observations of 51 Peg, obtaining results in excellent agreement
with previously measured orbital parameters due to its planetary
companion. We have also obtained measurements of a RV stable star,
$\eta$ Cas, showing that we can reach a precision of $7.9\ms$ over
seven days, using a simple and inexpensive prototype. The rms residuals
match the expected photon limited errors for the instrument, given
the fringe visibilities obtained, and show that the precision we
are able to obtain with \ET~is becoming comparable with current
traditional echelle techniques. For comparison, rms scatters
obtained previously for 51 Peg have been
$13\ms$\citep{51PegDiscovery}, $5.2\ms$ \citep{51PegConfirmation},
and $11.8\ms$ \citep{51peg_orbit}.

%%%%%%%%%%%%%%%%%%%%%%%%%%%%%%%%%%%%%%%%%%%%%%%%%%%%%%%%

\acknowledgments The authors are grateful to Richard Green, Skip
Andree, Daryl Wilmarth and the KPNO staff for their generous
support and advice, and to Dominique Naef for very helpful input. The authors are also grateful to Stuart Shaklan, Michael Shao and Chas Beichman
for their encouragement and support, and Bill Cochran, Larry Ramsey and Eric Feigelson for many useful discussions. This work is supported
by the National Science Foundation with grant AST-0243090, the
Penn State Eberly College of Science and JPL.
J. V. E. and S. M. acknowledge travel support from KPNO; S. M. acknowledges the JPL Michelson Fellowship funded by NASA.

%%%%%%%%%%%%%%%%%%%%%%%%%%%%%%%%%%%%%%%%%%%%%%%%%%%%%%%%%%%

%%%%%%%%%%%%%%%%%%%%%%%%%%%%%%%%%%%%%%%%%%%%%%%%%%%%%%%%%%

%%%%%%%
\clearpage
\begin{deluxetable}{rcc}
\tablecaption{RV measurements for 51 Peg \label{tab:51peg}}
\tablehead{\colhead{JD} & \colhead{Velocity} & \colhead{Error} \\
  \colhead{$-2450000$} & \colhead{$(\ms)$} & \colhead{$(\ms)$} }

\startdata

$ 2500.8665$ & $      35.4$ & $       7.1$ \\
$ 2500.8879$ & $      46.8$ & $       6.8$ \\
$ 2500.9094$ & $      29.4$ & $       6.8$ \\

$ 2501.9077$ & $      18.1$ & $       6.8$ \\
$ 2501.9263$ & $      30.8$ & $       7.0$ \\
$ 2501.9441$ & $      29.9$ & $       7.0$ \\

$ 2502.9392$ & $     -78.1$ & $       8.4$ \\
$ 2502.9573$ & $     -61.6$ & $       8.3$ \\
$ 2502.9751$ & $     -57.9$ & $       7.8$ \\

$ 2503.9250$ & $     -50.9$ & $       6.7$ \\
$ 2503.9480$ & $     -57.9$ & $       6.8$ \\
$ 2503.9692$ & $     -27.6$ & $       6.6$ \\

$ 2504.8960$ & $      44.2$ & $       6.5$ \\
$ 2504.9177$ & $      28.1$ & $       6.6$ \\
$ 2504.9392$ & $      15.4$ & $       6.8$

\enddata
\end{deluxetable}
%%%%%%%%

%%%%%%%%%%%%%%
\clearpage
\begin{deluxetable}{rcc}
\tablecaption{RV measurements for $\eta$ Cas \label{tab:EtaCas}}
\tablehead{\colhead{JD} & \colhead{Velocity} & \colhead{Error} \\
  \colhead{$-2450000$} & \colhead{$(\ms)$} & \colhead{$(\ms)$} }
\startdata
$ 2498.8318$ & $       6.4$ & $       5.7$ \\
$ 2498.8545$ & $       8.3$ & $       5.1$ \\
$ 2498.8823$ & $     -11.2$ & $       5.4$ \\
$ 2498.8936$ & $      10.9$ & $       5.3$ \\

$ 2499.7839$ & $       6.8$ & $       6.3$ \\
$ 2499.7952$ & $       2.9$ & $       6.0$ \\
$ 2499.8064$ & $      -3.6$ & $       5.8$ \\
$ 2499.8174$ & $       1.7$ & $       5.8$ \\
$ 2499.8284$ & $       1.4$ & $       5.9$ \\

$ 2501.8701$ & $      -5.9$ & $       5.3$ \\
$ 2501.8777$ & $      -9.6$ & $       5.4$ \\
$ 2501.8850$ & $     -15.3$ & $       5.5$ \\
$ 2501.8926$ & $       7.7$ & $       5.4$ \\

$ 2502.8528$ & $      -0.0$ & $       4.9$ \\
$ 2502.8643$ & $      -1.7$ & $       4.9$ \\
$ 2502.8752$ & $       5.4$ & $       5.1$ \\
$ 2502.8860$ & $     -10.8$ & $       5.4$ \\

$ 2504.8093$ & $      -1.4$ & $       5.7$ \\
$ 2504.8201$ & $     -10.1$ & $       6.4$ \\
$ 2504.8311$ & $       2.5$ & $       5.4$ \\
$ 2504.8401$ & $      11.7$ & $       5.9$

\enddata
\end{deluxetable}

%%%%%%%
\clearpage
\begin{deluxetable}{lcc}
\tablecaption{\label{tbl:errors}Mean photon limited error
estimation}
\tablehead{\colhead{Component} & \colhead{Star} & \colhead{Iodine} \\
     \colhead{ } & \colhead{($\ms$)} & \colhead{($\ms$)} }
\tablecolumns{3}

\startdata \sidehead{51 Peg}
Templates & 4.7  & 1.2 \\
Data &      6.7 & 7.2 \\
Combined & \multicolumn{2}{c}{11.0} \\

\sidehead{$\eta$ Cas}
Templates & 4.2 & 1.2 \\
Data &      5.3 & 4.3 \\
Combined & \multicolumn{2}{c}{8.1}

\enddata
\end{deluxetable}
%%%%%%%%

\clearpage

%%%%%%%
%%%%%%%%

%%%%%%%
%%%%%%%%

%%%%%%%
%%%%%%%%


\begin{thebibliography}{}

\bibitem[Baranne et al.(1996)]{Baranne1996} Baranne, A.~et al.\ 
1996, \aaps, 119, 373 
\bibitem[Butler et al.(1996)]{Butler1996} Butler, R.~P., Marcy, 
G.~W., Williams, E., McCarthy, C., Dosanjh, P., \& Vogt, S.~S.\ 1996, 
\pasp, 108, 500 
\bibitem[Edser \& Butler(1898)]{EdserButler}Edser, E. \& Butler, C. P.  1898, Phil. Mag., 46, 207
\bibitem[Erskine \& Ge(2000)]{ErskineGe} Erskine, D.~J.~\& Ge,
J.\ 2000, ASP Conf.~Ser.~195: Imaging the Universe in Three Dimensions, 501
\bibitem[Ge(2002)]{GeTheory} Ge, J.\ 2002, \apjl, 571, L165
\bibitem[Ge, Erskine, \& Rushford(2002)]{GeErskine} Ge, J.,
Erskine, D.~J., \& Rushford, M.\ 2002, \pasp, 114, 1016
\bibitem[Ge et al.(2003a)]{Ge2003a} Ge, J., van Eyken, J.~C., 
Mahadevan, S., DeWitt, C., Ramsey, L.~W., Shaklan, S.~B., \& Pan, X.\ 2003a, 
\procspie, 4838, 503 
\bibitem[Ge et al.(2003b)]{Ge2003b} Ge, J., Mahadevan, S., van Eyken, J.,
DeWitt, C., \& Shaklan, S. 2003b, in ASP Conf. Ser. 294, Scientific Frontier in Research in Extrasolar
Planets, ed. Deming, D., \& Seager, S. (San Fransico: ASP), 573
\bibitem[Lawson(2000)]{SummerSchool}Lawson, P. R.  2000, in Principles of Long Baseline Stellar Interferometry, ed. P. R. Lawson (Pasadena: JPL Publications), 113
\bibitem[Mahadevan et al.(2003)]{SuvrathSPIE} Mahadevan, S., Ge, J.,
van Eyken, J. C., DeWitt, \& Shacklan, S.  2003, \procspie, in press
\bibitem[Marcy et al.(1997)]{51PegConfirmation} Marcy, G.~W., Butler,
R.~P., Williams, E., Bildsten, L., Graham, J.~R., Ghez, A.~M., \& Jernigan,
J.~G.\ 1997, \apj, 481, 926
\bibitem[Mayor \& Queloz(1995)]{51PegDiscovery} Mayor, M.~\& Queloz,
D.\ 1995, \nat, 378, 355
\bibitem[Mosser, Maillard, \& Bouchy(2003)]{Mosser2003} Mosser, 
B., Maillard, J., \& Bouchy, F.\ 2003, \pasp, 115, 990 
\bibitem[Naef et al.(2003)]{51peg_orbit}Naef, D., Mayor, M., Beuzit,
J. L., Perrier, C., Queloz, D., Sivan, J. P., Udry, S.  2003, \aap, in press
\bibitem[van Eyken et al.(2003)]{vaneykenSPIE} van Eyken, J. C., Ge,
J., Mahadevan., S., DeWitt, C., \& Ren, D.  2003, \procspie, in press

\end{thebibliography}
\end{document}